\documentclass[preprint,amsmath,amssymb,nofootinbib,acs]{revtex4}

\usepackage{amsfonts}
\usepackage{amssymb}
\usepackage{latexsym}
\usepackage{graphicx}
\usepackage{longtable}
\usepackage{rotating}
\usepackage[active]{srcltx}
\newcommand{\al}{\alpha}

\newcommand{\De}{\Delta}

\newcommand{\rar}{\rightarrow}
\newcommand{\lrar}{\leftrightarrow}

\begin{document}

\title{The Ground State of the ${\rm H}_3^+$ Molecular Ion: Physics Behind}

\author{A.V.~Turbiner}
\email{turbiner@nucleares.unam.mx}

\author{J.C. Lopez Vieyra}
\email{vieyra@nucleares.unam.mx}

\affiliation{Instituto de Ciencias Nucleares, Universidad Nacional
Aut\'onoma de M\'exico, Apartado Postal 70-543, 04510 M\'exico,
D.F., Mexico}

\date{\today}

\begin{abstract}
Five physics mechanisms of interaction leading to the binding of the ${\rm H}_3^+$ molecular ion are identified. They are realized in a form of variational trial functions and their respective total energies are calculated. Each of them provides subsequently the most accurate approximation for the  Born-Oppenheimer (BO) ground state energy among (two-three-seven)-parametric trial functions being correspondingly, H$_2$-molecule plus proton (two variational parameters), H$_2^+$-ion plus H-atom (three variational parameters) and generalized Guillemin-Zener (seven variational parameters). These trial functions are chosen following a criterion of physical adequacy. They include the electronic correlation in the exponential form $\sim\exp{(\gamma r_{12})}$, where $\gamma$ is a variational parameter. Superpositions of two different mechanisms of binding are investigated and a particular one, which is a generalized Guillemin-Zener plus H$_2$-molecule plus proton (ten variational parameters), provides the total energy at the equilibrium of $E=-1.3432$\ a.u. The superposition of three mechanisms: generalized Guillemin-Zener plus (H$_2$ -molecule plus proton) plus (H$_2^+$ -ion plus H) (fourteen parameters) leads to the total energy which deviates from the best known BO energy to $\sim 0.0004$\ a.u., {\it it reproduces two-three significant digits in exact, non-BO total energy}. In general, our variational energy agrees in two-three-four significant digits with the most accurate results available at present as well as major expectation values.
\end{abstract}
%


\maketitle

\section{Introduction}

We had learned recently from Takeshi Oka about a fundamental importance of the two-electron hydrogenic molecular ion ${\rm H}_3^+$ in Physics, Chemistry and Astronomy \cite{Oka:2006},
in particular, that is a system of the lowest total energy among those made from one-two-three protons and electrons. The ion ${\rm H}_3^+$ is one of the most abundant chemical compounds in the Universe being a major proton donor in chemical reactions in interstellar space. Experimentally, the
${\rm H}_3^+$ was discovered by J.J.~Thomson in 1912~\cite{Thomson:1912}. On the other hand, being theoretical physicists we are not aware about any popular quantum-mechanical textbook nor a book on spectral theory where the ${\rm H}_3^+$ molecular ion is even mentioned. Even though the ${\rm H}_3^+$ as the simplest three-center problem plays the same role of a fundamental object as the hydrogen and helium atom being the simplest one-center problems, and ${\rm H}_2^+$ and
H$_2$ being the simplest two-center problems. Needless to say that such a role implies to use
${\rm H}_3^+$ as a test ground for different theoretical approaches.

In general, the Coulomb system $(3p,2e)$ is very difficult for theoretical studies. Many theoretical methods were developed to study low-lying quantum states of this system. In particular, it became clear very quickly that interelectron correlation is of great importance and it has to be included into the variational trial function explicitly to assure a faster convergence. This conclusion is similar to the one drawn by Hylleraas~\cite{Hylleraas:1930} for the ${\rm He}$ atom, and by James and Coolidge~\cite{JC:1933} for the ${\rm H}_2$ molecule. Usually, the interelectron correlation was written in a monomial form $r_{12}^n$ (Hylleraas~\cite{Hylleraas:1930} - James-Coolidge~\cite{JC:1933} form) or $\exp(-\al r_{12}^2)$ (Gaussian form, see e.g. Ref.~\cite{Kutzelnigg:2006} and references therein). However, quite recently it was shown (see \cite{Korobov:2000} and references therein) for the Helium atom that the use of the exponential form $\exp(\gamma r_{12})$ dramatically improves the convergence and leads, in fact, to the most accurate results for the ground state energy at present. Furthermore the similar use of the exponential correlation  $\exp(\gamma r_{12})$ for the ${\rm H}_2$ molecule allowed to construct the most accurate trial function among the few-parametric trial functions~\cite{TG:2007}. A hint of why this $r_{12}$-dependence leads to the fast convergent results was given in \cite{TG:2007}.

In 2006, the Royal Society discussion meeting on ${\rm H}_3^+$ took place in London, UK (see \cite{Oka:2006}) and in 2012 the Royal Society Theo Murphy Meeting on ${\rm H}_3^+$ was held where different properties of the  ${\rm H}_3^+$ ion and, in particular, various theoretical approaches to a study of the ${\rm H}_3^+$ ion were presented (see \cite{Oka:2012}). It is worth mentioning that the benchmark results for the potential energy surface of ${\rm H}_3^+$  were reported by Adamowicz - Pavanello \cite{AP:2012}.

The goal of this contribution is twofold: (i) to study a physics of binding, and based on that (ii) to propose a maximally simple, compact and easy-to-handle trial function with few (non)linear parameters which would lead to a highly accurate Born-Oppenheimer ground state energy and major expectation values. We always assume the two-Coulomb charge effective interaction in a presence of other Coulomb charges to be modeled by the wave function $e^{-\al r}$, where $\al$ is a parameter (which is a lowest Coulomb orbital for the case of two charges of opposite signs), since a corresponding potential reproduces a Coulomb singularity at small distances and vanishes at large distances. Perhaps, it is worth noting that we are not aware of previous studies of the ${\rm H}_3^+$ ion with trial functions involving $r_{12}$ in an exponential form with a single exception \cite{TGL:2007} where the ${\rm H}_3^+$ in a linear configuration in a magnetic field was explored. This paper can be considered as a continuation of a study of different physics mechanisms of binding started in \cite{LTM:2011}, where two possible mechanisms of binding and their superposition were explored (not dominant ones).

In this paper atomic units ($\hbar=e=m_e=1$) are used throughout.

\section{The ${\rm H}_3^+$ ion in the Born-Oppenheimer approximation}

\begin{figure}[tb]
\begin{center}
   \includegraphics*[width=4.in,angle=0.0]{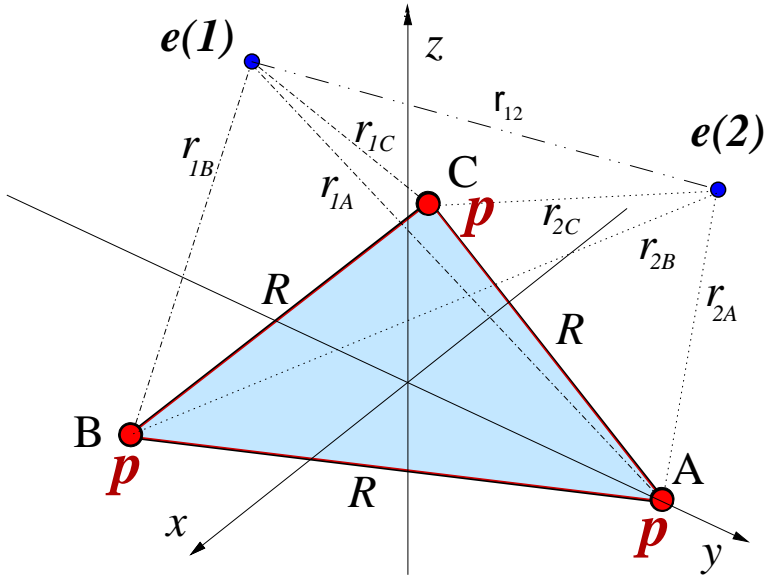}
    \caption{\label{fig1} Geometrical setting  for the hydrogen molecular ion ${\rm H}_3^+$ in equilateral triangular configuration.  The three protons are located on the $x$-$y$ plane
    forming an equilateral triangle with the origin of coordinates located at the geometrical center  (circumcenter) of the triangle.}
\end{center}
\end{figure}

The Hamiltonian which describes the ion ${\rm H}_3^+$  under the assumption that the protons are infinitely massive (the Born-Oppenheimer approximation of zero order) and located at the vertices of an equilateral triangle of side $R$
(see Fig.~\ref{fig1} for the geometrical setting and notations), is written as
follows:
\begin{equation}
\label{H}
  {\cal H}\ =\sum_{j=1}^2 \frac{{\hat {\mathbf p}_{j}}^2}{2}\ -
  \ \sum_{\buildrel{{j}=1,2}\over{\kappa =A,B,C}}
  \frac{1}{r_{{j},\kappa}}
  \ +\ \frac{1}{{\sf r}_{12}}\ +\ \frac{3}{R}  \ ,
\end{equation}
where ${\hat {\mathbf p}_{j}}=-i \nabla_{j}$ is the 3-vector
of the momentum of the ${j}$th electron, the index $\kappa$ runs
over protons $A$, $B$ and $C$, $r_{{j},\kappa}$ is the distance
between the ${j}$th electron and the $\kappa$th proton, ${\sf r}_{12}$ is the
interelectron distance, and  $R$ is the interproton distance.

It is a well established fact that the ground state of the ${\rm H}_3^+$ molecular ion
is $1 {}^1 A_1^\prime$, an electronic  spin-singlet state, with the three protons forming
an equilateral triangle in the totally symmetric  representation $A_1^\prime$  of a
$D_{3h}$ point symmetry \cite{Tennyson:1995}. Thus, the ground state electronic wavefunction
should be symmetric under permutations of the three indistinguishable protons. This ground state
is the major focus of the present study.

It is worth  mentioning that the best theoretical value at the moment for the Born-Oppenheimer ground state energy is \hbox{$E=-1.343 835 625 02$\,a.u.} \cite{Pavanello:2009} obtained with a basis of 1000 explicitly correlated spherical Gaussian functions with shifted centers. This value surpasses the previous record
$E=-1.343 835 624$\,a.u. by Cencek {\it et al.} which was obtained by using
explicitly correlated Gaussian functions \cite{Cencek:1995}.

\section{Variational method}

We use the variational procedure to explore the problem. The recipe for
choosing the trial function is based on arguments of physical relevance, e.g. the trial
function should support the symmetries of the system, has to reproduce the Coulomb
singularities and the asymptotic behavior at large distances adequately
(see, e.g. \cite{turbinervar, turbinervar1, Turbiner:2006}). To the problem
finding a bound state of $(3p, 2e)$ giving rise to ${\rm H}_3^+$ molecular ion
we follow the description given in \cite{LTM:2011}.

In practice, the use of such trial functions based on physical adequacy
implies the convergence of a special form of the perturbation theory where
the variational energy is the sum of the first two terms of the perturbation series.
Let us recall the essentials of this perturbation theory (for details, see \cite{turbinervar, turbinervar1, Turbiner:2006}). Let us assume that the original Hamiltonian has a form ${\cal H}=-\De + V$, where $\De$ is the Laplacian. As a first step we choose a trial function $\psi^{(trial)}$ (normalized to one) and find a potential for which such a trial function $\psi^{(trial)}$ is an exact eigenfunction, {\it i.e.}
$V_{trial}={\De \psi^{(trial)}}/{\psi^{(trial)}}$, with energy $E_{trial}=0$. In a pure formal way we can construct a Hamiltonian ${\cal H}_{trial} = -\Delta + V_{trial}$ such that ${\cal H}_{trial} \psi^{(trial)}=0$. It can be easily shown that the
variational energy
\[
 E_{var}= \langle\psi^{(trial)}|{\cal H}|\psi^{(trial)}\rangle
\]
is nothing but the first two terms in the perturbation theory where
the unperturbed problem is given by ${\cal H}_{trial}$ and the
perturbation is the deviation of the original potential $V$ from the
trial potential $V_{trial}$, namely, $V_{perturbation}=V-V_{trial}$.
Eventually, we arrive at the formula
\begin{equation}
    E_{var}= E_{trial} + E_1 (V_{perturbation})\ ,
\end{equation}
here $E_1
(V_{perturbation})=\langle\psi^{(trial)}|V_{perturbation}|\psi^{(trial)}\rangle$
is the first energy correction in the perturbation theory, where
the unperturbed potential is $V_{trial}$. It is worth noting that if the
trial function is the Hartree-Fock function, the resulting
perturbation theory is nothing but the Moeller-Plesset perturbation
theory (see, e.g. \cite{Levine}, Section 15.18)\footnote{It is
worth noting that the question about convergence of the Moeller-Plesset perturbation
theory is not settled yet \cite{MP}}.

One of the criteria of convergence of the perturbation theory in
$V_{perturbation}=V-V_{trial}$ is a requirement that the ratio
$|{V_{perturbation}}/{V}|$ should not grow when $r$ tends to
infinity in any direction. If this ratio is bounded by a constant it
should be less than one. In fact, it is a condition that the
perturbation potential is subordinate with respect to the
unperturbed potential. The value of this constant controls the rate of
convergence - a smaller value of this constant leads to faster
convergence \cite{turbinervar1}. Hence, the above condition underlines the
importance of the large-range behavior of the trial functions. In
the physics language the above requirement means that the phenomenon
of the Dyson's instability should not occur (for a discussion see
\cite{turbinervar}) \footnote{It is worth noting that this procedure
for a selection of the trial function was applied successfully to a
study of one-two-electron molecular systems in a magnetic field leading
to highly accurate results. Many of these results are the most
accurate at present (see \cite{Turbiner:2006} and \cite{Turbiner:2010}).}.

In order to make the above-mentioned criteria of convergence concrete for the problem
of ${\rm H}_3^+$ let us introduce the exponential representation of the spin-singlet
ground state function
\begin{equation}
\label{exp}
    \Psi (x) \ =\ (1+ P_{12}) \sum_{{\rm perm}\{ A,B,C\}}
 e^{ -\varphi(r_{1A} ,r_{1B}, r_{1C}, r_{2A}, r_{2B}, r_{2C}, \gamma r_{12})}\ ,
\end{equation}
where $\varphi (r_1, r_2) \equiv \varphi(r_{1A} ,r_{1B}, r_{1C}, r_{2A}, r_{2B}, r_{2C}, \gamma r_{12})$ is unknown phase, and $P_{12}$ is the permutation operator which interchanges 
electrons $(1 \lrar 2)$. If (\ref{exp}) is substituted to the Schroedinger equation a non-linear
equation of a type of the multidimensional Riccati equation will occur. The analysis of the
behavior of $\varphi$ at large distances along any direction in six-dimensional physical space
${\bf R^6}$ leads to the asymptotics of the form
\begin{equation}
\label{rl}
    \varphi \rar a r + O(\log r)\ , \quad r \rar \infty \ ,
\end{equation}
where the constant $a$ can depend on the chosen direction, while the behavior near a Coulomb singularity $r_C$ of the potential is
\begin{equation}
\label{rC}
    \varphi \rar b r + O(r^2)\ , \quad r \rar r_C \ ,
\end{equation}
where $b$ makes sense of a cusp parameter (residue at the Coulomb singularity).
Hence, a trial function with a phase $\varphi_{trial}$ which grows linearly at large distances guarantees a convergence of the perturbation theory where the chosen trial function is considered as zero approximation.
Furthermore, for such a choice the ratio $|\varphi_{trial}/\varphi_{exact}|$ is bounded. Actually, this
property implies that relative accuracies in energy and in an expectation value may not be very much
different.

There are the different ways to include  electronic correlation explicitly in the
trial wave function for two-electron problems. We mention three major approaches
(see e.g. \cite{Klopper:2006}): the linear in $r_{12}$, the gaussian $\exp(-\al r_{12}^2)$
and exponential $\exp(\gamma r_{12})$ terms. Among them, the only factor
$\exp(\gamma r_{12})$ fulfills the physical adequacy requirements for a trial
function described above and may lead to a convergent perturbation theory.

\section{Physics mechanisms of binding}

As an illustration to what follows let us describe physics mechanisms of binding
of the simplest two-center, one-electron molecular system ${\rm H}_2^+$. There are
three of them: (i) a coherent interaction of electron with two centers, (ii)
incoherent interaction of electron with two centers, and (iii) a general, screened
interaction of electron with two centers. A simplest realization of these three mechanisms leads to the celebrated Heitler-London function (the product of two lowest, equally screened Coulomb orbitals, the electron "sees" two centers simultaneously), Hund-Mulliken function (the sum of two lowest, equally screened Coulomb orbitals, the electron "sees" one center only, it corresponds to the interaction H-atom with proton), and Gulliemin-Zener function (the symmetrized product of two lowest, unequally screened Coulomb orbitals, the electron "sees" both centers but differently). A linear superposition of these three functions gives the accuracy of order of $10^{-5}$ for any separation (see e.g. \cite{Turbiner:2006}).

Following the  guidelines of Section III and the convergence requirement of the perturbation theory, the most general trial function for the ground state made out of individual exponentials is of the form:
\begin{equation}
\label{ansatzG}
 \psi_{G}\ =\ (1+ P_{12}) \sum_{{\rm perm}\{ A,B,C\}}
 e^{
 -\al_1 r_{1A}-\al_2 r_{1B} -\al_3 r_{1C} -\al_4 r_{2A}
 -\al_5 r_{2B} -\al_6 r_{2C}
 + \gamma r_{12} }\ ,
 \end{equation}
see Fig.1 for notations, where the sum runs over the permutations of the identical protons $A,B,C$ ($S_3$ symmetry), and $P_{12}$ is the operator which interchanges electrons $(1 \lrar 2)$. The variational parameters consist of non-linear parameters $\al_{1-6}$ and $\gamma$ which characterize the (anti)screening of the Coulomb charges. The interproton distance $R$, see Fig.1 is kept fixed and all there distances are chosen to be equal, thus, protons form equilateral triangle, $R=R_{eq}=1.65$ a.u. [10-11]
\footnote{
Almost all previous calculations were carried out for the same value of $R_{eq}$. It is justified by the fact that the minimum of the potential curve in $R$ is very flat being characterized by a small curvature. Hence, for a given accuracy in energy the accuracy of localizing the position of the minimum is much lower (see Table I)
}. The function (\ref{ansatzG}) is a symmetrized product of $1s$ Slater type orbitals multiplied by the exponential correlation factor $e^{\gamma r_{12}}$.

Calculations of the variational energy were performed using the minimization package
MINUIT from CERN-LIB. Six-dimensional integrals which appear in the functional of energy were calculated numerically using a "state-of-the-art" dynamical partitioning procedure. The calculations were made in double-cylindrical coordinate system: $(x_1, y_1,z_1,x_2,y_2,z_2) \rar (\rho_1, \phi_1,z_1,\rho_2,\phi_2,z_2)$.
As the first step the infinite domain of integration was reduced to a finite, compact domain in a form of two 3D cylinders in such a way that a contribution from the complement should be two orders of magnitude smaller than the requested absolute accuracy of integration. As the second step the domain of integration was subdivided into 972 up to 6534 subdomains following the profile of the integrand, in particular, separating out
the domains with sharp changes of the integrand, the domains of the large gradients. Then each
subdomain was integrated separately in a parallel manner with controlled absolute/relative accuracy
(for details, see e.g. \cite{Turbiner:2006}).
A realization of the routine required a lot of attention and care. During the minimization process both the domain and partitioning were permanently controlled and adjusted. Numerical integration of every subdomain was done with a relative accuracy \hbox{of $\sim 10^{-3} - 10^{-7}$} depending on its complexity and relative contribution using an adaptive routine based on an algorithm by Genz and Malik~\cite{GenzMalik}. This is an adaptive multidimensional integration routine (CUBATURE) of vector-valued integrands over hypercubes
written by Steven G. Johnson with a vectorized prototype prepared by Dmitry Turbiner,
http://ab-initio.mit.edu/wiki/index.php/Cubature.
Parallelization was implemented using the MPI library MPICH. The code was written as the hybrid:
some parts were in FORTRAN (in particular, minimization and parallelization) and other parts were
in C (in particular, integration). We consider as an important future task to rewrite this code in C++.

Computations were performed on a Linux cluster with 48 Xeon processors at 2.67\,GHz each, and 12Gb total RAM plus an extra processor serving as the master node. The complete minimization process for every Ansatz took from 100 (single Ansatz, Eqs. (6-11), see below)) up to 2000 (triple mechanism Ansatz, Eq. (16), see below) hours of aggregated wall clock time. A single integration was about from 1 up to 15 minutes. With optimal values of parameters, it took about 60 minutes (wall clock time) to compute a variational energy (calculating two integrals) with relative accuracy $10^{-7}$.

\subsection{Single mechanism}

A goal of subsection is to identify some possible single mechanisms of interactions which occur in the system of three static protons and two electrons, $(3p 2e)$.

\subsubsection{Coherent interaction}

One of the most natural mechanisms of interaction is when each electron "sees" all charged centers simultaneously, coherently interacting with them. Such a mechanism has to be dominant at small internuclear distances in comparison with a characteristic length of the electron-proton interaction which seems to be of the order of 1 a.u. = 1 Bohr radius. A simplest realization of this mechanism in a wavefunction is given by a product of Coulomb orbitals
\begin{equation}
\label{ansatz-coh}
 \psi_{coh}\ =\  e^{-\al_1 (r_{1A}+ r_{1B} + r_{1C} + r_{2A} + r_{2B} + r_{2C})
      + \gamma r_{12} }\ ,
\end{equation}
(Ansatz 1) which is a generalization of the Heitler-London function for the case of the H$_2^+$ molecular ion or the H$_2$ molecule; it contains two variational parameters $\al_1, \gamma$ both of them having a meaning of charge (anti)screening. Variational energy as well as the corresponding values of the parameters are presented in Table~\ref{table1}. It is already a striking fact that even at $R_{eq} = 1.65$ a.u. which is essentially larger than the Bohr radius the function (\ref{ansatz-coh}) leads to a binding of the $(3p 2e)$ system with sufficiently large ionization energy $\sim 0.12$~a.u. (H$_3^+ \rar$~H$_2$~+~$p$) \footnote{In order to calculate the ionization energy we use the H$_2$ total energy -1.1745 a.u. (rounded) found in \cite{SH:2006}}.

\begin{table}
\centering
\begin{tabular}{|l|l|l|ccccccc|}
\hline
& Trial Function &\ $E$ (Ry)& $\al_1$ & $\al_2$  & $\al_3$ & $\al_4$ & $\al_5$ &
$\al_6$ & $\gamma$
\\
 \hline \hline
{\it 1.} &\ Coherent      &\ -2.583   &  0.5711  & 0.5711  &  0.5711 & 0.5711  &  0.5711
& 0.5711  & 0.29783   \\
{\it 2.} &\ Incoherent    &\ -2.595   &  1.3557  & 0       &  0      & 0       &  1.3557
& 0       &-0.06136   \\
{\it 3.} &\ $H^- + 2p$    &\ -2.148   &  1.3284  & 0       &  0      & 1.3284  &  0
& 0       & 0.25244   \\
{\it 4.} &\ $H_2   + p$   &\ -2.634   &  0.9422  & 0.9422  &  0      & 0.9422  &  0.9422
& 0       & 0.43977   \\
{\it 5.} &\ $H_2^+ + H$   &\ -2.650   &  0.9114  & 0.9114  &  0      & 0       &  0
& 1.2499  & 0.04359   \\
{\it 6.} &\ Generic       &\ -2.680 7\ &\ -0.00353 & 0.18548 & 1.4245 & 1.0471  & 0.15082
& 0.58912 & 0.21632
\\
 \hline \hline
\end{tabular}
\caption{\label{table1}
  The ground state energy of the $(3p 2e)$ system in different mechanisms of binding realized by Ansatz 1 - 6.}
\end{table}

\subsubsection{Incoherent interaction}

Another natural mechanism of interaction is when each electron "sees" a single charge center, thus, realizing an incoherent interaction. In fact, it corresponds to the interaction of two hydrogen atoms and proton, H + H + $p$. Such a mechanism has to be dominant at large internuclear distances in comparison to a characteristic length of the electron-proton interaction of the order of 1 a.u. = 1 Bohr radius. A simplest realization of this mechanism is given by
\begin{equation}
\label{ansatz-incoh}
 \psi_{incoh}\ =\  (1+ P_{12}) \sum_{{\rm perm}\{ A,B,C\}}e^{-\al_1 (r_{1A} + r_{2B})
 + \gamma \ r_{12} }\ ,
\end{equation}
(Ansatz 2) which is a generalization of the Hund-Mulliken function for H$_2^+$ ion or H$_2$ molecule, it contains two variational parameters $\al_1, \gamma$ having a meaning of charge (anti)screening. Variational energy as well as the corresponding values of the non-linear parameters $\al_1, \gamma$ are presented in Table~\ref{table1}. At $R_{eq} = 1.65$ a.u. which is essentially larger than the Bohr radius, the function (\ref{ansatz-incoh}) leads to a binding of the $(3p 2e)$ system with sufficiently large ionization energy $\sim 0.12$\ a.u. ${}^4$. The parameter $\gamma$ takes a negative value indicating the effective attraction of the electrons in this mechanism.

\subsubsection{(almost)Incoherent interaction}

This mechanism of interaction appears when two electrons "see" the same charge center interacting coherently, then, the interaction with two other charged centers appears as a result of the exchange interaction. In fact, it corresponds to the interaction of negative hydrogen ion with two protons, H$^-$ + $2 p$. Such a mechanism has to be important at large internuclear distances comparing to a characteristic length of the electron-proton interaction of the order of 1 a.u. = 1 Bohr radius. It has to be suppressed due to a small probability of the coherent interaction of two electrons with same proton. A simplest realization of this mechanism is given by
\begin{equation}
\label{ansatz-incoh-a}
 \psi_{(a)incoh}\ =\  (1+ P_{12}) \sum_{{\rm perm}\{ A,B,C\}}e^{-\al_1 (r_{1A} + r_{2A}) + \gamma r_{12} }\ ,
\end{equation}
(we call it the Ansatz 3). Perhaps, it is not surprising that such a trial function does not lead to a binding (see Table I).

\subsubsection{H$_2$-molecule + proton interaction}

In this case both electrons "see" two protons forming H$_2$ molecule which interacts with proton. Though the interproton distance $R_{eq} = 1.65$ a.u. for H$_3^+$ is larger than the equilibrium distance for the H$_2$ molecule, which is equal to $\sim 1.4$ a.u., it can be described by the Heitler-London function for the H$_2$ molecule. Eventually, as a realization of this mechanism we choose
\begin{equation}
\label{ansatz-h2}
 \psi_{H_2}\ =\  (1+ P_{12}) \sum_{{\rm perm}\{ A,B,C\}}e^{-\al_1 (r_{1A} + r_{1B}+r_{2A} + r_{2B}) + \gamma r_{12} }\ ,
\end{equation}
(Ansatz 4) which contains two variational parameters $\al_1, \gamma$ having a meaning of charge (anti)screening. Variational energy as well as the corresponding values of the non-linear parameters $\al_1, \gamma$ are presented in Table~\ref{table1}. The function (\ref{ansatz-h2}) leads to a binding with sufficiently large ionization energy $\sim 0.15$\ a.u. It is worth noting that an attempt to improve a description of the H$_2$ molecule by replacing the Heitler-London function by the Guillemin-Zener one
\[
e^{-\al_1 (r_{1A} + r_{1B}+r_{2A} + r_{2B})} \rar e^{-\al_{1,1} (r_{1A} + r_{1B})
 - \al_{1,2} (r_{2A} + r_{2B})}
\]
leads to unessential improvement in ionization energy $\sim 0.16$ a.u. ${}^4$.

\subsubsection{H$_2^+$-ion + H-atom interaction}

In this case one electron "sees" two protons forming H$_2^+$ ion and the second electron "sees" the third proton forming H-atom. The interproton distance $R_{eq} = 1.65$ a.u. for H$_3^+$ is smaller than the equilibrium distance for the H$_2^+$ ion, which is equal to $\sim 2.$ a.u., H$_2^+$ contribution can be described by the Heitler-London function for the H$_2^+$ ion. Eventually, as a realization of this mechanism we choose
\begin{equation}
\label{ansatz-h2+h}
 \psi_{H_2^+ H}\ =\  (1+ P_{12}) \sum_{{\rm perm}\{ A,B,C\}}e^{-\al_{1} (r_{1A} + r_{1B}) -
 \al_{6} r_{2C}  + \gamma r_{12} }\ ,
\end{equation}
(Ansatz 5) which contains three variational parameters $\al_{1}, \al_{6}, \gamma$ having a meaning of charge (anti)screening. This function can be considered as a generalization of the celebrated Guillemin-Zener function written for the H$_2^+$-ion and the H$_2$-molecule.
Variational energy as well as the corresponding values of the parameters are presented in Table~\ref{table1}. The function (\ref{ansatz-h2+h}) leads to a binding with sufficiently large ionization energy $\sim 0.14$\ a.u. ${}^4$ which is nevertheless smaller than one for the Ansatz {\it 4.} The parameter $\gamma$ is close to zero indicating effectively the small repulsion of the electrons.

\subsubsection{Generic interaction}

Naturally, a general mechanism of interaction is related to interaction of an electron with a proton (or another electron) with its particular (anti)screening due to the presence of other charges. A simplest realization of this mechanism is given by the function (\ref{ansatzG}) (see above). The variational parameters consist of non-linear parameters $\al_{1-6}$ and $\gamma$ which characterize the (anti)screening of the Coulomb charges. Appropriate degeneration of these parameters allow us to reproduce different mechanisms of interaction {\it 1.-5}. Variational energy as well as the corresponding values of the parameters are presented in Table~\ref{table1}.
The function (\ref{ansatzG}) leads to a binding with quite high ionization energy $\sim 0.17$\ a.u.${}^4$.

\subsection{Mixed mechanisms}

In previous subsection we identified possible single mechanisms of interactions which occur in the system of three static protons and two electrons, $(3p 2e)$. Based on the values of ionization (binding) energies these mechanisms can be classified coming from the highest ionization energy at {\it Generic interaction} mechanism
(Ansatz {\it 6.}) followed by the mechanism {\it H$_2^+$-ion + H-atom interaction} (Ansatz {\it 5.}) and then {\it H$_2$-molecule + proton interaction} (Ansatz {\it 4.}) . The goal of this subsection is to consider different superpositions of single mechanisms. The mechanism {\it (almost)Incoherent interaction} (H$^-$ + $2p$, Ansatz {\it 3.}) is excluded as one giving no binding to the $(3p 2e)$ system. In all mixed mechanisms the {\it Generic interaction} mechanism will be always present as one giving the highest ionization energy among single mechanisms.

\begin{sidewaystable}
\centering
\begin{tabular}{|l|c|ccccccc|ccc|}
\hline
  & $E$ (Ry)   & $\al_1$  & $\al_2$ & $\al_3$ & $\al_4$ & $\al_5$ &
$\al_6$ & $\gamma$ & $A$ & $\tilde\al $ & $\tilde\gamma$  \\
\hline \hline
 \qquad Generic Ansatz (6) &\ -2.680 7 &\ -0.00353 & 0.18548 & 1.4245  & 1.0471  &
0.15082 & 0.58912 & 0.21632\ & - & - & - \\
\hline \hline
 \ (6) $\oplus$ (7)       &\ -2.683 2  &\ -0.00294 & 0.21022 & 1.3849 & 1.0199 & 0.17103
& 0.59084  & 0.26044    \ &\ -0.51154   &  0.59589  &\  0.86229     \\
\hline \hline
 \ (6) $\oplus$ (8)       &\ -2.683 5  &\ -0.02840& 0.19525& 1.4780& 1.06860& 0.12045 &
0.52108  & 0.24008      \ &\ -0.03987   &  1.71230  &\  0.69991     \\
\hline \hline
 \ (6) $\oplus$ (11) (H$_2^+$ + H) &\ -2.685 4 &\ -0.05699 & 0.22600  & 1.40290  & 1.04611&
0.17386 &0.55836 & 0.14867 \ &\ -1.52370  &  1.50704  &\ -1.21701  \\
&&&&&&&&&                    &  1.10855  &\\
\hline \hline
 \ (6) $\oplus$ (10) (H$_2$)    &\ -2.686 4  &\ 0.00515 &  0.19795 & 1.4043  &  1.03200 &
0.13466  &  0.59350 &  0.13686 \ &\ -2.34310 &  0.99547 &\ -0.46046\\
\hline \hline
 \ (6) $\oplus$ (10) (H$_2$) $\oplus$ (11) (H${}_2^+$ + H)\  &\ -2.686 852\ \ &\ 0.00040 & 0.23034
&1.42450 &
1.00418 & 0.12283 & 0.56933 & 0.13409\ &
 \ -2.34400 & 0.94900 & -0.55946 \\
&&&&&&&&& \ \ 0.16243 & 1.72960  & -0.87202 \\
&&&&&&&&&             & 0.066457 &          \\
\hline \hline
 \qquad Best value \cite{Pavanello:2009} (rounded) &\ -2.687 671 &  &  &  &  &
 & &  &  &  & \\
\hline\hline
\end{tabular}
\caption{\label{table2}
  The ground state energy of ${\rm H}_3^+$ at \hbox{$R_{eq}=1.65$\,a.u.}
and the non-linear variational parameters
in $[a.u.]^{-1}$ corresponding to the trial function (\ref{ansatzG}) and
different degenerations of it. For the case
(6) + H${}_2^+$ + H, the $\al$ parameter in the second row corresponds to the
effective charge for the H-atom part.}
\end{sidewaystable}

\subsubsection{Generic interaction plus coherent interaction}

Let us consider a superposition of the {\it Generic interaction} (Ansatz {\it 6.})
and {\it Coherent interaction} (Ansatz {\it 1.}) (see \cite{LTM:2011})
\begin{equation}
\label{61}
   \psi_{61} \ =\ \ \psi_{G}\ +\ A_{coh} \psi_{coh}\ .
\end{equation}
Formally, the trial function $\psi_{61}$ depends on ten  parameters (one linear and 9 non-linear).
The results are presented in Table~\ref{table2}. It is interesting to note that the non-linear parameters are almost unchanged in comparison to the parameters of the individual Ansatze {\it 1.} and {\it 6.} with a single exception of the parameter $\tilde \gamma$. If it is compared with the Ansatz {\it 6.} this Ansatz increases the ionization energy to $ \sim 0.001$ \ a.u.

\subsubsection{Generic interaction plus incoherent interaction}

Let us consider a superposition of the {\it Generic interaction} (Ansatz {\it 6.})
and {\it Incoherent interaction} (Ansatz {\it 2.})
\begin{equation}
\label{62}
   \psi_{62} \ =\ \ \psi_{G}\ +\ A_{incoh} \psi_{incoh}\ .
\end{equation}
Formally, the trial function $\psi_{62}$ depends on ten  parameters (one linear and 9 non-linear).
The results are presented in Table~\ref{table2}. It is interesting to note that the non-linear parameters of the Ansatz {\it 6.} in (\ref{62}) are almost unchanged in comparison to the parameters of the individual Ansatz {\it 6.} with a single exception of the parameter $\al_1$ while the parameters of the Ansatz {\it 2.} are changed significantly. If it is compared with the Ansatz {\it 6.} this Ansatz increases the ionization energy to $ \sim 0.002$ \ a.u.

\subsubsection{Generic interaction plus H$_2$-molecule + proton interaction}

Let us consider a superposition of the {\it Generic interaction} (Ansatz {\it 6.})
and {\it H$_2$-molecule + proton interaction} (Ansatz {\it 4.})
\begin{equation}
\label{64}
   \psi_{64} \ =\ \ \psi_{G}\ +\ A_{H_2} \psi_{H_2}\ .
\end{equation}
Formally, the trial function $\psi_{64}$ depends on ten  parameters (one linear and 9 non-linear).
The results are presented in Table~\ref{table2}. It is interesting to note that the non-linear parameters of the Ansatz {\it 6.} in (\ref{64}) are almost unchanged in comparison to the parameters of the individual Ansatz {\it 6.} with a single exception of the parameter $\gamma$ while the parameters of the Ansatz {\it 3.} are changed insignificantly. If it is compared with the Ansatz {\it 6.} this Ansatz increases the ionization energy to $ \sim 0.003$ \ a.u. In comparison
with the most accurate total energy this Ansatz reproduces 3 figures.

\subsubsection{Generic interaction plus H$_2^+$-ion + H-atom interaction}

Let us consider a superposition of the {\it Generic interaction} (Ansatz {\it 6.})
and {\it H$_2^+$-ion + H-atom interaction} (Ansatz {\it 5.})
\begin{equation}
\label{65}
   \psi_{65} \ =\ \ \psi_{G}\ +\ A_{H_2^+ + H} \psi_{H_2^+ + H}\ .
\end{equation}
Formally, the trial function $\psi_{65}$ depends on eleven parameters (one linear and 10 non-linear).
The results are presented in Table~\ref{table2}. It is interesting to note that the non-linear parameters of the Ansatz {\it 6.}
in (\ref{65}) are almost unchanged in comparison to the parameters of the individual Ansatz {\it 6.} with a single exception of the parameter $\gamma$ while the parameters of the Ansatz {\it 4.} are changed significantly, especially the parameter $\gamma$. If it is compared with the Ansatz {\it 6.} this Ansatz increases the ionization energy to $ \sim 0.002$ \ a.u. In comparison
with the most accurate total energy this Ansatz reproduces 3 figures.

\subsubsection{Generic interaction plus H$_2$-molecule + proton interaction plus H$_2^+$-ion + H-atom interaction}

Following the analysis of the variational total energies found in single mechanisms of binding of the H$_3^+$-ion we arrive at the hierarchy: {\it Generic interaction} (Ansatz {\it 6.}) goes first, then
{\it H$_2^+$-ion + H-atom interaction} (Ansatz {\it 5.}) is the second and {\it H$_2$-molecule + proton interaction} (Ansatz {\it 4.}) as the third. In turn, the analysis of the variational total energies found in mixed mechanisms of binding involving two mechanisms we arrive again at a conclusion that above-mentioned three mechanisms of binding play a dominant role. Thus, a natural move is to consider a superposition of these three mechanisms of binding:
the {\it Generic interaction} (Ansatz {\it 6.}), {\it H$_2^+$-ion + H-atom interaction} (Ansatz {\it 5.}) and {\it H$_2$-molecule + proton interaction} (Ansatz {\it 4.}),
\begin{equation}
\label{645}
   \psi_{645} \ =\ \ \psi_{G}\ +\ A_{H_2}\ \psi_{H_2}\ +\ A_{H_2^+ + H}\ \psi_{H_2^+ + H}\ .
\end{equation}
The trial function $\psi_{645}$ depends on fourteen parameters (two linear and 12 non-linear).
The results are presented in Table~\ref{table2}. It is interesting to note that remarkably
the non-linear parameters of the Ansatz {\it 6.} in (\ref{645}) are almost unchanged in comparison
to the parameters of the individual Ansatz {\it 6.} itself as well as in Ansatz (\ref{65}) and Ansatz (\ref{64}) with a single exception of the parameter $\gamma$ which nevertheless remains almost unchanged for Ansatz (\ref{65}), Ansatz (\ref{64}) and Ansatz (\ref{645}). The non-linear parameters of the Ansatz {\it 5.} are slightly changed in the Ansatz (\ref{65}) and Ansatz (\ref{645}). We consider this as an indication that the {\it Generic interaction} is described adequately by the Ansatz {\it 6.} (\ref{ansatzG}) as well as the {\it H$_2$-molecule + proton interaction} is described adequately by the Ansatz {\it 4.} (\ref{ansatz-h2}). The non-linear parameters of the Ansatz {\it 5.} are changed significantly, especially the parameter $\gamma$, comparing to ones in the Ansatz (\ref{64}) and
Ansatz (\ref{645}). Also the non-linear parameters of the Ansatz {\it 4.} in the Ansatz (\ref{64}) and Ansatz (\ref{645}) are very much different. We consider this as an indication that the mechanism of the {\it H$_2^+$-ion + H interaction} is described inadequately by the Ansatz {\it 4.}
(\ref{ansatz-h2+h}). Perhaps, for H$_2^+$-ion part the Heitler-London function for H$_2^+$ should be replaced by more adequate Guillemin-Zener function
\[
     e^{-\al_{4,1} (r_{1A} + r_{1B})} \rar e^{-\al_{4,1}^{(1)} r_{1A} - \al_{4,1}^{(2)} r_{1B}}
\]
(see (\ref{ansatz-h2+h})). It will be done elsewhere.

If the Ansatz (\ref{645}) is compared with the Ansatz {\it 6.} this Ansatz leads to increase in the ionization energy to $ \sim 0.003$ \ a.u. Making a comparison of the obtained variational total energy with the most accurate total energy it appears that this Ansatz provides a deviation from the best known energy which is equal to $\sim 0.0004$\ a.u. (see Table~\ref{table3}).

\section{Results}

In Table~\ref{table3} we present the results for the ground state energy at interproton
equilibrium distance of the ${\rm H}_3^+$ molecular ion obtained previously using different methods.
It can be seen that the Born-Oppenheimer ground state energy obtained using the trial function
(\ref{645}) is the most accurate (the lowest) energy obtained with a few parametric functions.
In particular, the trial function (\ref{645}) gives a lower energy than the energies obtained
with the explicitly correlated functions based on both Gaussians in $r_{12}$ \cite{Salmon:1973}
and linear in $r_{12}$ \cite{Klopper:1993}, when a relatively small number of terms with
non-linear parameters is involved. The trial function (\ref{645}) is more accurate than
all(!) traditional CI calculations which were performed before 1990 (see \cite{Anderson:1992}) even
including the one with the largest set of 100 configurations \cite{DubenII:1971, Kawaoka:1971}
\footnote{for a list of 42 calculations of the ground state energy of ${\rm H}_3^+$ in the period
1938-1992 see Ref.\cite{Anderson:1992}, for a list of selected {\it ab-initio} calculations till
1995, see \cite{Kutzelnigg:2006}} and as well as [32], however, being worse than the large GG
calculation \cite{Alexander:1990} (700 configurations). In above CI calculations no explicit
correlation was included.
The trial function (\ref{645}) is more accurate than some CI calculations with explicitly involved
$r_{12}$ dependence \cite{Preiskorn:1982} (36 configurations) and \cite{Preiskorn:1984} (192 configurations) but gets worse when (much) larger number of configurations is included, see \cite{Urdaneta:1988}-\cite{Frye:1990}. In general, the 14-parametric trial function (\ref{645}) gives the eighth best result among known in literature so far (see Table~\ref{table3}) being incomparably simpler with respect to all known trial functions. It is worth noting that the 7-parametric trial function (\ref{ansatzG}) \cite{LTM:2011} gives the thirteenth best result among known in literature (see Table~\ref{table3}).

The list of major expectation values obtained using different trial functions (\ref{ansatzG}), (12) - (16) and its comparison with results of other calculations is given in Table~\ref{table4}.
A reasonable agreement for the expectation values is observed. In particular, for the expectation values $\langle 1/r_{1 A}\rangle$, $\langle x^2\rangle$, $\langle z^2\rangle$ and $\langle r^2\rangle$ agreement within $\sim$ 1\% with ours and all other calculations is observed, including ones obtained in the large CISD-R12 calculations \cite{Klopper:1993}.
Also, for the expectation value of $\langle 1/r_{12}\rangle$ we have an agreement in the first significant digit with other calculations, and we are also in close agreement to the value obtained with the correlated Gaussian (unrestricted) wavefunction with 15 terms \cite{Salmon:1973}. For the expectation values $\langle 1/r_{1 A}\rangle$ and $\langle z^2 \rangle$ we observe agreement with other calculations in 3 and 2 significant digits, respectively. These facts seem to indicate that the presented expectation values are very accurate, corroborating the quality of the trial functions (\ref{ansatzG}), (12)-(16) which gives 2-3 s.d. correctly. It is worth noting that since there are no criteria about accuracy of the obtained expectation values, we can only comment about the agreement within our results based on different trial functions and those obtained by other approaches. Since the exact phase of the ground state eigenfunction is uniformly approximated by the phases the trial functions (6), (12) - (16) we guess that the relative accuracy reached in variational energy has to be similar to the relative accuracy in an expectation value. It implies that the expectation value based on (6), (12) - (15) should provide two significant digits correctly, while (16) should give three significant digits (see Table~\ref{table4}). Following
this assumption we can state that $<r_{12}>=1.9(9)$ a.u., $<1/r_{12}>=0.62(6)$ $(a.u.)^{-1}$ (hence, in Ref.\cite{DubenII:1971} a single figure was found correctly), $<1/r_{1A}>=0.85(5)$ $(a.u.)^{-1}$
(it agrees in 2 figures with all previous calculations, see Table~\ref{table4}),  $<x_1^2>=0.76(1)$ $(a.u.)^{2}$ (it agrees in 2 figures with all previous calculations, see Table~\ref{table4}),
$<z_1^2>=0.54(2)$ $(a.u.)^{2}$ (it agrees in 2 figures with all previous calculations
except (12) and (14), see Table~\ref{table4}),
$<r_1^2>=2.0(6)$ $(a.u.)^{2}$ (it agrees in 2 figures with all previous calculations, see Table~\ref{table4})

\begin{table}
\begin{tabular}{|l|l|l|l|}
\hline
\hspace{10pt} $E$ (a.u.) & $R$ (a.u.) &  \hspace{40pt} method & reference
\\ \hline \hline
 -1.339 7  & 1.66  &  CI-GTO,  $5s2p$ basis set  & \cite{Ciszmadia:1970}
(1970)\\ \hline
 -1.306 29  & 1.65     &   GG, 3 terms, 5 non-linear params   &
   \cite{Salmon:1973} (1973) \\
 -1.327 25  & 1.65     &   GG, 6 terms, 7 non-linear params   &
     \\
 -1.331 47  & 1.65     &   GG, 10 terms, 9 non-linear params   &
      \\
 -1.332 29  & 1.65     &   GG, 15 terms, 11 non-linear params   &
      \\  \hline
 -1.334 382 &  1.65  &  R12, $10s$ basis set  &  \cite{Klopper:1993}
(1993)  \\
 -1.334 632 &  1.65  &  R12, $30s$ basis set  &     \\ \hline
 -1.340 34  &  1.65  &  7-Parametric Trial Function (\ref{ansatzG}), 1 config &
\cite{LTM:2011} (2011) \\
 \hline
 -1.340 50   & 1.6406 &   CI -GTO, 48 configs   &   \cite{DubenII:1971}
(1971)\\ \hline
 -1.340 5   & 1.65   &   CI -STO, 100 configs  &   \cite{Kawaoka:1971}
(1971)\\ \hline
-1.342 72   & 1.65041 & CI-GTO, 108 terms & \cite{Burton:1985} (1985)\\
 -1.342 784 & 1.6504 & CI-GTO,  $8s3p1d/[6s3p1d]$ basis set&
\cite{Dykstra:1978} (1978)\\
-1.343 40   & 1.6504  & CI-GTO,  $10s4p2d$ basis set (with additional terms) &
\cite{Meyer:1986}
(1986)\\
\hline
-1.342 03 & 1.6504  & CI with r12, 36 configs & \cite{Preiskorn:1982}
(1982)\\
 -1.343 422 & 1.6504  & CI with r12, 192 configs & \cite{Preiskorn:1984}
(1984)\\
 -1.343 426 & 1.65  & 14-parametric Trial function (\ref{645}), 3 configs  & present \\
-1.343 500 &  1.6405  & R12, $13s3p/[10s2p]$ basis set &
\cite{Urdaneta:1988} (1988)\\
-1.343 828  & 1.65  & CI with r12,  $13s5p3d$ basis set. &
\cite{Frye:1990} (1990)\\ \hline
 -1.343 835  & 1.65   &  R12, $30s20p12d9f$ basis set  &
\cite{Klopper:1993} (1993)  \\ \hline
 -1.343 35  & 1.65    &   GG, 15 terms, 135 non-linear params   &
   \cite{Salmon:1973} (1973) \\
-1.343 822  & 1.65   & GG, 700 terms& \cite{Alexander:1990} (1990)\\
 -1.343 835 624  &  1.65  &  GG,  600 terms  &  \cite{Cencek:1995} (1995)
\\ \hline
 -1.343 835 625 02 & 1.65 & ECSG, 1000 terms & \cite{Pavanello:2009} (2009)\\
\hline
\end{tabular}
\caption{\label{table3}
 A selection of the calculations for the Born-Oppenheimer ground state
energy at equilibrium distance of ${\rm H}_3^+$. Record calculations of
Ref.~\cite{Pavanello:2009} (2009) and Ref.\cite{Cencek:1995} (1995). CI
denotes Configuration Interaction, STO - Slater Type Orbitals, GTO -
Gaussian Type Orbitals, GG - correlated Gaussians (Gaussian Geminals),
R12 - the CI calculation augmented by terms linear in $r_{12}$,
CI with r12 - the CI calculation augmented by polynomials in $r_{12}$,
ECSG - Explicitly Correlated Spherical Gaussian functions.
Present results: Trial Function (\ref{645}) with fourteen parameters
optimized.}
\end{table}
\begin{table}[htdp]{
\begin{center}
\begin{tabular}{|l|lc|c|}
\hline
Expectation Value & \multicolumn{2}{|c|}{\hspace{10pt}Trial Function} &
Others\\
\hline
$\langle r_{12}\rangle$                  & 2.0032   & (\ref{ansatzG}) &
\\[-5pt]
                                         & 1.9931   & (\ref{61})      &  \\
                                         & 1.9927   & (\ref{62})      &  \\
                                         & 1.9831   & (\ref{65})      &  \\
                                         & 1.9868   & (\ref{64})      &  \\
                                         & 1.9864   & (\ref{645})     &
\\ \hline
$\langle 1/r_{12}\rangle$                & 0.6315   & (\ref{ansatzG}) &
0.59549${}^a$ \\[-5pt]
                                         & 0.6302   & (\ref{61})      &
0.62636${}^c$ \\
                                         & 0.6293   & (\ref{62})      &  \\
                                         & 0.6273   & (\ref{65})      &  \\
                                         & 0.6280   & (\ref{64})      &  \\
                                         & 0.6261   & (\ref{645})     &
\\ \hline
$\langle 1/r_{1 A}\rangle$               & 0.8548  & (\ref{ansatzG}) &
0.85519${}^c$ \\[-5pt]
                                         & 0.8549  & (\ref{61})      &
0.8553${}^e$ \\
                                         & 0.8553  & (\ref{62})      &  \\
                                         & 0.8555  & (\ref{65})      &  \\
                                         & 0.8556  & (\ref{64})      &  \\
                                         & 0.8553  & (\ref{645})     &  \\
\hline
\end{tabular}
\end{center}
\caption{\label{table4}
   Expectation values (in a.u.) for the ${\rm H}_3^+$ ion in its ground
state obtained
   with the trial functions (\ref{ansatzG}) and (\ref{61}-\ref{645}) .
Corresponding results obtained with other methods are displayed for
comparison. The coordinates $x,y,z$  and $r$ are measured from the
center of the equilateral triangle formed by protons.
   ${}^a$ Ref.\cite{DubenII:1971} CI-48;
   ${}^b$ CI wavefuncion (I) in Ref.\cite{Kawaoka:1971};
   ${}^c$ Correlated Gaussian (unrestricted) wavefunction with 15 terms
in Ref.\cite{Salmon:1973};
   ${}^d$ CI wavefunction in Ref.\cite{Carney:1974};
   ${}^e$ CISD-R12 wavefunction with the $10s8p6d4f$  basis set in
Ref.\cite{Klopper:1993}.
}
} 
\end{table}%
\addtocounter{table}{-1}
\begin{table}[htdp]{
\begin{center}
\begin{tabular}{|l|lc|c|}
\hline
Expectation Value & \multicolumn{2}{|c|}{\hspace{10pt}Trial Function} &
Others\\
\hline
$\langle x_1^2\rangle =\langle y_1^2\rangle$
                                         & 0.7711   & (\ref{ansatzG}) &
0.75429${}^a$ \\[-5pt]
                                         & 0.7666   & (\ref{61})      &
0.7595${}^b$  \\
                                         & 0.7610   & (\ref{62})      &
0.75913${}^c$ \\
                                         & 0.7588   & (\ref{65})      &
0.75968${}^d$ \\
                                         & 0.7609   & (\ref{64})      &
0.7605${}^e$  \\
                                         & 0.7613   & (\ref{645})     & \\
\hline
$\langle z_1^2\rangle$                     & 0.5399   & (\ref{ansatzG})  &
0.54129${}^a$ \\[-5pt]
                                         & 0.5337   & (\ref{61})       &
0.5451${}^b$  \\
                                         & 0.5420   & (\ref{62})       &
0.54085${}^c$ \\
                                         & 0.5375   & (\ref{65})       &
0.54179${}^d$ \\
                                         & 0.5348   & (\ref{64})       &
0.5396${}^e$  \\
                                         & 0.5420   & (\ref{645})      &
\\ \hline
$\langle r_1^2\rangle$                     & 2.0822   & (\ref{ansatzG}) &
2.04988${}^a$ \\[-5pt]
                                         & 2.0669   & (\ref{61})      &
2.0640${}^b$  \\
                                         & 2.0640   & (\ref{62})      &
2.05911${}^c$ \\
                                         & 2.0550   & (\ref{65})      &
2.06114${}^d$ \\
                                         & 2.0565   & (\ref{64})      &   \\
                                         & 2.0646   & (\ref{645})     &   \\
\hline
\end{tabular}
\end{center}
\caption{{\it Continued.} Expectation values (in a.u.) for the ${\rm H}_3^+$ ion in
its ground state obtained with the trial functions (\ref{ansatzG}) and (\ref{61}-\ref{645}) .
Corresponding results obtained with other methods are displayed for
comparison. The coordinates $x,y,z$  and $r$ are measured from the
center of the equilateral triangle formed by protons.
   ${}^a$ Ref.\cite{DubenII:1971} CI-48;
   ${}^b$ CI wavefuncion (I) in Ref.\cite{Kawaoka:1971};
   ${}^c$ Correlated Gaussian (unrestricted) wavefunction with 15 terms
in Ref.\cite{Salmon:1973};
   ${}^d$ CI wavefunction in Ref.\cite{Carney:1974};
   ${}^e$ CISD-R12 wavefunction with the $10s8p6d4f$  basis set in
Ref.\cite{Klopper:1993}.}
} 
\end{table}%

\section{Conclusion}

We identified five mechanisms of interaction leading to the binding in the Coulomb problem of two electrons in the field of three fixed charged centers $(2e 3p)$ in equilibrium geometry - the charged centers form the equilateral triangle with side 1.65 a.u. Among these five mechanisms three
lead to a strongest binding: {\it H$_2$-molecule plus proton}, {\it H$_2^+$ion plus H-atom} and generic. Each mechanism of interaction is easily modeled by a certain two-three-seven-parametric wavefunction, respectively, which is taken as a variational trial function. Corresponding variational energies are certainly the lowest ones among two-three and seven parametric trial functions, respectively (see Table~\ref{table3}).

In particular, a simple and compact 7-parametric variational trial function (\ref{ansatzG}) (generic mechanism) already provides a surprisingly accurate Born-Oppenheimer energy for the ground state of such a complicated molecular system as ${\rm H}_3^+$. The minimum energy is found to be $E=-1.340 34$\,a.u. at an equilibrium interproton distance $R=1.65$\,a.u.  This result for the energy is the most accurate among the values obtained with several parametric trial functions of GG type \cite{Salmon:1973} (Gaussian in $r_{12}$), when a relatively small number of terms and non-linear parameters are involved, see Table~\ref{table3}. In particular, it is more accurate than the energies obtained with the explicitly correlated approaches of Ref.\cite{Klopper:1993}.

It seems natural to advance by considering a superposition of three dominant mechanisms of binding, like it was done in \cite{TG:2007} for H$_2$-molecule,
\begin{equation}
\label{psi3}
    \Psi = \psi_{G}\ +\ A_{H_2} \psi_{H_2}\ +\ A_{H_2^+ + H} \psi_{H_2^+ + H}\ ,
\end{equation}
as well as a superposition of two dominant mechanisms of binding. It immediately
gives an essential improvement in the energy (see Table~\ref{table2}). In particular, the function (16) (a superposition of three dominant mechanisms), which contains fourteen variational parameters, allows us to get a more accurate result for the energy than one obtained by Preiskorn-Woznicki \cite{Preiskorn:1984} in CI with $r_{12}$ included using 192 configurations. Our result is slightly worse than one by \cite{Urdaneta:1988} in CI with $r_{12}$ included, using $13s3p/[10s2p]$ basis set.

Each of the trial functions $\psi_{G}\,, \  \psi_{H_2}\,,\ \psi_{H_2^+ + H}$ in (16) are neither normalized nor orthogonal. We can make them orthogonal using the Gram-Schmidt procedure and then normalize defining the orthonormal set as
\begin{eqnarray*}
\widetilde{\psi}_G
         &=& {\cal N}_{G} \, \psi_{G} \, \,, \\
\widetilde{\psi}_{H_2^++H}
         &=& {\cal N}_{H_2^++H} \left(  \psi_{H_2^++H}  + \al \ {\psi}_G
                                  \right)\,, \\
\widetilde{\psi}_{H_2}
         &=& {\cal N}_{H_2} \left(  \psi_{H_2} + \beta_1 {\psi}_G +
\beta_2  {\psi}_{H_2^++H}
\right)
            \,.
\end{eqnarray*}
where $\al, \beta_{1,2}$ are the parameters of mixing and ${\cal N}$'s are normalization factors.
In terms of these functions the expression (\ref{psi3}) takes the form
\begin{equation}
\label{psi3tilde}
 \Psi = {\tilde A}_G\,\widetilde{\psi}_G  +
        {\tilde A}_{H_2^++H}\,\widetilde{\psi}_{H_2^++H}  +
        {\tilde A}_{H_2}\,\widetilde{\psi}_{H_2}\,.
\end{equation}
In the case of minimal energy at equilibrium configuration (see Table II)
we find $\al =-0.1111$ and $\beta_1 = -0.0107\ ,\ \beta_2 = -0.2553$ while
${\tilde A}_G=3.763160$, ${\tilde A}_{H_2^++H}=-0.130344$, and ${\tilde A}_{H_2}=-0.180807$.
Thus, the relative contribution of each mechanism is given by norm expansion
\begin{equation}
 \int |\Psi|^2 dV = 1 =  {r_G} +  r_{{\rm H}_2^+{\rm H}} + r_{{\rm H}_2}\ +\ \ldots     \,,
\end{equation}
where $\Psi$ is given by (\ref{psi3tilde}),
\[
  r_{G}                  \approx  \frac{{{\tilde A}_{G}}^2}{{\cal N}^2}         =
0.99650  \,, \quad
  r_{{\rm H}_2^+{\rm H}} \approx  \frac{{{\tilde A}_{H_2^++H}}^2}{{{\cal N}^2}} =
0.00120  \,, \quad
  r_{{\rm H}_2}          \approx  \frac{{{\tilde A}_{H_2}}^2}{{{\cal N}^2}}     =
0.00230  \,.
\]
and ${\cal N}$ is normalization factor for the ground state wavefunction. Thus, one can draw a conclusion that the dominant mechanism of the binding is generic one while the particular mechanisms (H$_2$ + p) and (H$_2^+$-ion + H-atom) give small contributions. It implies that H$_3^+$ at equilibrium is true many-body system which can not be split into interacting sub-systems.

\begin{figure}[tb]
\begin{center}
   \includegraphics*[width=4in,angle=-90.0]{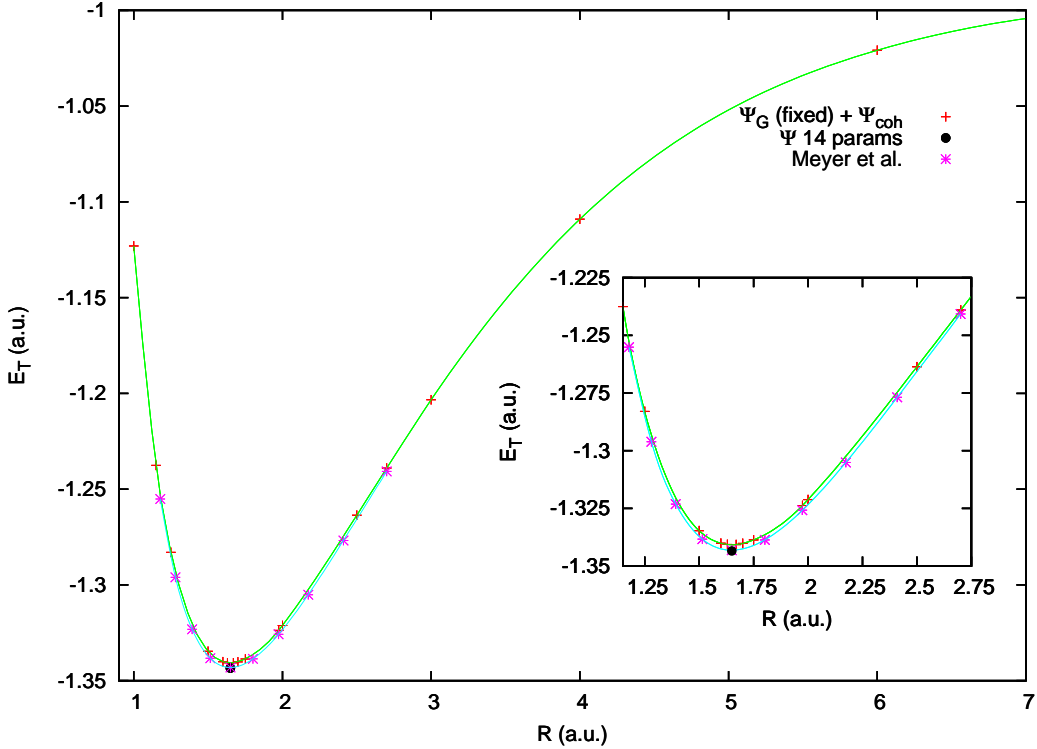}
    \caption{\label{figx} Potential Energy Curve for the ${\rm H}_3^+$ ion in triangular equilateral configuration as a function of the triangle size $R$ obtained with the 7+3 parameter Ansatz (\ref{61}) (see \cite{LTM:2011}). Results by Meyer {\it et al.} obtained using CI $10s,4p,2d$ GTO basis \cite{Meyer:1986} are shown for comparison. The energy of (16) at equilibrium is marked by bullet}
\end{center}
\end{figure}

The obtained energy (16) is among the eight most accurate variational results ever
calculated so far but using the incomparably simpler trial function. The major expectation values in Table \ref{table4} are gradually changed with the moves from one Ansatz to another seemingly demonstrating a convergence. It is clearly seen that in any expectation value (as well as in BO energy) two significant digits remain stable, they do not depend on the Ansatz considered. Thus, it can be stated that two significant digits in any expectation value are calculated reliably. The relation $\langle x_1^2\rangle =\langle y_1^2\rangle$, which holds for $D_{3h}$ symmetric configurations, was first mentioned in \cite{Kawaoka:1971}.
Without doubt, the trial function (\ref{ansatzG}) can be used to study the potential energy surface.
It seems that two significant digits are guaranteed. As an illustration in Fig.~2 the potential curve $E$ vs $R$ is presented. It is based on a linear superposition of Ansatz (\ref{ansatzG}) and (\ref{ansatz-coh}).

It is worth emphasizing that the main virtue of functions (\ref{ansatzG}), (\ref{645}) is their compactness. A similar analysis of binding can be carried out for low-lying states and corresponding wavefunctions can be written. In particular, the function (\ref{ansatzG}) can be easily modified for a study of spin-triplet states, as well as the low-lying states with non-vanishing magnetic quantum number. A generalization to more-than-two electron molecular systems seems also straightforward.

In order to conclude we must mention that the domain of applicability of the BO approximation of the zero order (infinitely massive particles of positive charge are assumed) is limited. This question can be thorough studied in the example of the two-center problem exploring the H$_2^+$, D$_2^+$, T$_2^+$ ions. We used the ground state energies and the expectation values for the internuclear distance obtained in the Lagrange-mesh method for two-center problem $E_{BO}, R_{eq}$ and the H$_2^+$, D$_2^+$, T$_2^+$ ions \cite{Horop:2012} with relative accuracy $\sim 10^{-12}$ (for energies) and $\sim 10^{-7}$ (for the internuclear distances) the fits yield
\[
    E(M,m_e) - E_{BO}\ =\ 0.468436\ \sqrt{\frac{m_e}{M}} \quad , \quad <R_{NN}(M,m_e)> - R_{eq}\ =\ 2.84964\ \sqrt{\frac{m_e}{M}}\ ,
\]
where $m_e$ is the electron mass and $M$ is the mass of a heavy particle. Hence, the finite-mass corrections begin to contribute to the 2-3rd significant digit. One way to include the finite-mass corrections is to introduce the zero energy state corresponding to the lowest energy state of nuclear motion. In this case the zero energy is equal to the BO ground state energy of the zero order (a minimum at the potential curve) plus the first vibrational energy. One of the ways to find the first vibrational state is to consider a harmonic approximation of the bottom of the potential curve. One can see that the zero energy coincides with the non-BO energy up to four-five significant digits. In order to get further improvement it is necessary to go beyond the harmonic approximation, also taking into account non-adiabatic corrections and write more accurate BO trial functions. It makes the overall analysis complicated. We think it is more relevant to consider an alternative way treating the problem solving the exact 3-body Schr\"odinger equation without any type of BO approximation similar to what was done for example in \cite{OB:2012}. In particular, it would allow to study the essential effects which do not exist in BO approximations like the electric quadrupole transitions.

Similar considerations can be performed for three-center problem finding $E_{BO}, R_{eq}$ and for
the H$_3^+$, D$_3^+$, T$_3^+$ ions. From non-BO calculations made in \cite{Bubin:2005} it can be seen that finite-mass corrections are essentially larger than those for the two-center case. They  contribute to the 2-3rd significant digits of the ground state energies of H$_3^+$, D$_3^+$, T$_3^+$ and the expectation value for internuclear distance,
\[
    E(M,m_e) - E_{BO}\ =\ 1.27232\ \sqrt{\frac{m_e}{M}} \quad , \quad <R_{NN}(M,m_e)> - R_{eq}\ =\ 4.21669\ \sqrt{\frac{m_e}{M}}\ ,
\]
where $m_e$ is the electron mass and $M$ is the mass of a heavy particle.
Taking into account the lowest vibrational energy found in \cite{Kutzelnigg:2006} the agreement between zero energy and non-BO ground state energy \cite{Bubin:2005} is improved up to 4th significant digit only. We would consider it as indication that the exact, 5-body Schroedinger equation has to be treated when going beyond the fifth significant digit in energy. The consideration presented in \cite{Bubin:2005} shows that this is feasible. It seems evident that the dominant mechanisms of interaction leading to binding found in the BO approximation will remain dominant beyond, in non-BO considerations. We think that in such an approach particularly the radiative transitions in the H$_3^+$ ion have to be studied.

\section*{Acknowledgement}

We are deeply indepted to Takeshi Oka who firstly brought our attention to the H$_3^+$ ion and then taught us (unconsciously) about physics related to H$_3^+$. We thank to A Alijah for the interest in this work and many useful discussions.

This work was supported in part by PAPIIT grant {\bf IN115709} and CONACyT grant {\bf 166189} (Mexico). The authors are deeply thankful to D. Turbiner (MIT - JPL NASA) for designing
the computer code prototype and for the creation of an optimal configuration of a 48-processor cluster used for this calculation.
The authors are also obliged to E. Palacios for his technical support with the cluster {\it Karen} where most of calculations were done.
The authors are grateful to the University Program FENOMEC (UNAM, Mexico) for a partial support.

\end{document}